\documentclass[aps,preprint]{revtex4}

\newcommand{\fft}[2]{{\frac{#1}{#2}}}
\newcommand{\ft}[2]{{\textstyle\frac{#1}{#2}}}
\renewcommand{\Re}{{\mathrm{Re}}}

\newcommand{\be}{\begin{equation}}
\newcommand{\ee}{\end{equation}}
\newcommand{\bea}{\begin{eqnarray}}
\newcommand{\eea}{\end{eqnarray}}
\begin{document}
\title{AdS$_5$ Black Holes with Fermionic Hair}
\date{December 2004}

\author{Benjamin A. Burrington}
\email{bburring@umich.edu}
\author{James T. Liu}
\email{jimliu@umich.edu}
\affiliation{Michigan Center for Theoretical Physics,
Randall Laboratory of Physics, The University of Michigan,
Ann Arbor, MI 48109-1120, USA}
\author{W. A. Sabra}
\email{ws00@aub.edu.lb}
\affiliation{Center for Advanced Mathematical Sciences (CAMS)}
\affiliation{Physics Department, American University of Beirut, Lebanon}

\begin{abstract}
The study of new BPS objects in AdS$_5$ has led to a deeper
understanding of AdS/CFT.  To help complete this picture, and to fully
explore the consequences of the supersymmetry algebra, it is also
important to obtain new solutions with bulk fermions turned on.  In this
paper we construct superpartners of the $1/2$~BPS black hole in AdS$_5$
using a natural set of fermion zero modes.  We demonstrate that these
superpartners, carrying fermionic hair, have conserved charges differing
from the original bosonic counterpart.  To do so, we find the $R$-charge
and dipole moment of the new system, as well as the mass and angular
momentum, defined through the boundary stress tensor.  The complete set
of superpartners fits nicely into a chiral representation of AdS$_5$
supersymmetry, and the spinning solutions have the expected gyromagnetic
ratio, $g=1$.
\end{abstract}

\preprint{MCTP-04-70}
\preprint{CAMS/04-04}
\preprint{hep-th/0412155}

\maketitle

\section{Introduction}
Since the recognition of their importance in connecting weakly coupled
to strongly coupled physics, BPS states have continued to play a
major r\^ole in fulfilling the promises of strong/weak coupling duality.
This is certainly evident today in the exploration of AdS/CFT, where a
weakly coupled gravity system in five dimensions is dual to four-dimensional
super-Yang Mills theory at strong 't~Hooft coupling.  In general, very
few direct comparisons may be made between states at weak coupling and
strong coupling.  After all, following a state from weak to strong coupling
involves the observation of more and more corrections, until finally the
perturbative description, valid at weak coupling, breaks down altogether.
In many cases, even the effective degrees of freedom are expected to change,
so that keeping track of individual states would not make sense.

On the other hand, the reason BPS states are useful is that, as shortened
representations of the supersymmetry algebra, they are protected against
corrections by supersymmetry.  Thus, in contrast with arbitrary states,
they may be traced between strong and weak coupling.  As such, they provide
a primary means for extracting information out of systems which involve
strong/weak coupling duality.  For example, there is currently much interest
in the 1/2~BPS excitations of AdS$_5\times S^5$ configurations.  From the
bulk point of view, these states have interpretation as either gravitational
ripples or giant gravitons.  These may be investigated either as classical
solutions of the supergravity equations or through the world-volume dynamics
of wrapped branes.  Furthermore, through duality, such states are also
associated with chiral primaries in the dual field theory.  It is precisely
the BPS nature of such excitations that allow such a rich connection to be
made between seemingly different objects such as branes, classical gravity
backgrounds and chiral primary operators.

In a supergravity context, extremal black holes are an obvious choice as
BPS objects to explore.  Such black holes have zero temperature and have
a natural correspondence with pure states in a quantum theory.  In this
case, like all states in a supersymmetric model, the extremal black holes
ought to form representations of the supersymmetry algebra.  In particular,
the bosonic black hole solution itself must also be related to superpartner
black holes carrying fermionic hair.  In fact, such superpartners may be
constructed by action of a finite supersymmetry transformation $\delta$
on the original solution, represented schematically as
\begin{equation}
\Phi\longrightarrow e^\delta\Phi=\Phi+\delta\Phi+\ft12\delta\delta\Phi+\cdots.
\label{eq:deltadelta}
\end{equation}
Here, $\Phi$ would be the metric, graviphoton or any other field in
the supergravity theory.  For Poincar\'e supergravity, a typical example
would be the extremal Reissner-Nordstrom solution with mass $=$ charge.
Clearly this coincides with the corresponding BPS condition $M=|Z|$ where
$Z$ is a central charge in the supersymmetry algebra.  In this case, exactly
half of the supersymmetries $\delta\Phi$ would vanish, namely those related
to the Killing spinors of the background.  On the other hand, the remaining
fermion zero mode spinors would generate non-trivial transformations,
demonstrating that the black hole lies in a shortened multiplet of
supersymmetry.

This construction of exact black hole superpartners was first carried out in
\cite{Aichelburg:1986wv} in the context of ungauged ${\cal N}=2$ supergravity
in four dimensions.  The method was also used in \cite{Duff:1996bs} to
examine the dipole moments and gyromagnetic ratios of 1/2 and 1/4~BPS
black holes in ungauged $D=4$, ${\cal N}=8$ supergravity, in
\cite{Duff:1998ef} to construct the fermionic partners of the supergravity
description of D0-branes in ten dimensions, and in
\cite{Balasubramanian:1998za} to construct the M2-brane multiplet in
eleven dimensional supergravity.  In general, for such extremal
objects in Poincar\'e supergravity, it may be explicitly demonstrated
that all superpartners have identical masses and charges as the original
black hole itself.  At some level, this is simply a kinematical consequence
of satisfying the supersymmetry algebra.  Nevertheless, it is reassuring
to see that semi-classical methods may be successfully applied to the
study of new backgrounds with fermionic hair.

In this paper, we show that the same techniques for generating
black hole superpartners in ungauged supergravities may also be applied
to the case of gauged (or anti-de Sitter) supergravities.  However, it is
important to note that the AdS superalgebra is different from the
Poincar\'e one.  In particular, masses (actually energies) and charges
of the superpartners are no longer identical, but are related according
to the AdS superalgebra.  Below, we construct explicit superpartners for
black holes in gauged $D=5$, ${\cal N}=2$ supergravity and go on to
calculate the masses and charges of the superpartners.  We verify
that the mass and charge shifts indeed follow the pattern required by
supersymmetry in AdS spaces.

The main purpose behind this construction of black hole superpartners
is to demonstrate that the fermion zero modes carry additional information
about BPS background in supergravity.  Although we work explicitly with
black holes, the techniques we use are applicable to any background with
partially broken supersymmetry, even those without horizons.  By fully
studying the BPS states and their partners, we may also hope to obtain
new methods for exploring the lowest non-BPS excitations as well.  While
the zero-mode construction fails for non-BPS black holes (since the
would-be zero modes are non-normalizable at the horizon), this obstacle
may potentially be overcome in geometries without horizons.

It is worth noting that obtaining a meaningful definition of mass and
angular momentum in AdS spaces involves some care.  While various
definitions have been provided, we use the holographic renormalization
method \cite{Henningson:1998gx,Balasubramanian:1999re,Emparan:1999pm},
which is natural in an AdS/CFT context.  Properties of the
five-dimensional black holes which we are interested in have recently
been examined in \cite{Liu:2004it,counter}.  There it was demonstrated
that a proper set of boundary counterterms was necessary to ensure the
validity of the BPS algebra on the boundary.

We begin in section~\ref{sec:bh} with a brief overview of ${\cal N}=2$
supergravity and very special geometry as well as the familiar 1/2~BPS
black hole solutions themselves.  In section \ref{sec:vary} we identify
convenient zero mode spinors and use these to modify the bosonic
background of section \ref{sec:bh}.  In section \ref{sec:charges} we
calculate the conserved charges of the new system, defining the mass and
angular momentum through the boundary stress tensor.  We conclude in
section \ref{sec:dis} by showing that these black holes fall into
shortened chiral representations of AdS$_5$ supersymmetry and find the
gyromagnetic ratio of the spinning superpartners to be $g=1$.

\section{\label{sec:bh}BPS black holes in five dimensions}

We are interested in $1/2$~BPS black hole solutions of gauged
${\cal N}=2$ supergravity in five dimensions coupled to $n$ vector
multiplets.  The bosonic fields in this model consist of the metric
$g_{\mu\nu}$, $n+1$ vectors $A_\mu^I$, and $n$ scalars $\phi^x$,
while the fermionic fields are the gravitino $\psi_\mu$ and $n$
gauginos $\lambda_x$.
The gauging of a $U(1)$ subgroup of the $SU(2)$ $R$-symmetry group
is achieved by introducing a linear combination of the $n+1$ vectors,
${\cal A}_\mu=V_IA^I_\mu$ where the $V_I$ are a set of constants.  Since
${\cal A}_\mu$ is what couples to $R$-charge, it will play a prominent
r\^ole in the supersymmetry analysis of section~\ref{sec:dis}.

This theory was constructed in \cite{sugra,gauged}, where particular
attention was paid to the notion of very special geometry.  We follow
the conventions of \cite{BCS,Behrndt:1998jd} and write the bosonic
action as
\begin{equation}
e^{-1}{\mathcal L}=
\fft12 R -\fft14 G_{IJ}F^{I}_{\mu \nu} F^{\mu \nu \, J}
-\fft12 g_{xy}(\phi)\partial_{\mu}\phi^x \partial^{\mu}\phi^y
-V(\phi)+\frac{e^{-1}}{48}
\epsilon^{\mu\nu\rho\sigma\lambda}
C_{IJK}F^{I}_{\mu \nu}F^{J}_{\rho\sigma} A^K_{\lambda},
\label{action}
\end{equation}
where we use a signature $(-,+,+,+,+)$.
The gauging of the $U(1)$ subgroup introduces a potential $V(\phi)$
which may be obtained from a superpotential $W(\phi)$ through the relation
\begin{equation}
V(\phi)=\fft12g^{xy}\fft{\partial W}{\partial\phi^x}\fft{\partial W}{\partial\phi^y}
-\fft23W^2,
\end{equation}
where
\begin{equation}
W(\phi)=3gV_IX^I.
\label{eq:spot}
\end{equation}

For very special geometry, the $n$-dimensional scalar manifold is obtained
by introducing $n+1$ scalar coordinates $X^I(\phi)$ along with the restriction
${\cal V}(X^I)=1$ where ${\cal V}$ is a homogeneous cubic
polynomial
\begin{equation}
{\cal V}=\fft16C_{IJK}X^IX^JX^K.
\label{eq:calv}
\end{equation}
In this case, we have
\be
G_{IJ}=-\fft12 \left. \left(\frac{\partial}{\partial X^I}
\frac{\partial}{\partial X^J}\ln{\cal V}\right)\right|_{{\cal V}=1},\qquad
g_{xy}=G_{IJ}\partial_x X^I \partial_y X^J,
\ee
where $\partial_x\equiv\partial/\partial\phi^x$.  We also find it convenient
to introduce $X_I=\fft16C_{IJK}X^JX^K$, so that $X^IX_I=1$ as well as
$X_I dX^I=X^IdX_I=0$, so long as we restrict ourselves to ${\cal V}=1$.

In addition to the bosonic sector, given by (\ref{action}), we will also
need the supersymmetry variations
\bea
\delta \psi_{\mu} &=& \left({\mathcal D}_{\mu} +
\ft{i}{8} X_I\left(\Gamma_{\mu}{}^{\nu
  \rho}-4\delta_{\mu}^{\nu}\Gamma^{\rho}\right)F_{\nu \rho}^I
  +\ft16W\Gamma_{\mu}\right)\epsilon,\nonumber\\
\delta \lambda_x &=& \left(\ft38 \partial_x X_I \Gamma^{\mu \nu}F^I_{\mu
  \nu} - \ft{i}{2}g_{xy}\Gamma^{\mu}\partial_{\mu}\phi^y + \ft{i}2
\partial_x W\right)\epsilon,
\label{eq:fsusy}
\eea
for the fermions, and
\bea
\delta g_{\mu \nu}&=&
\Re\left(\bar{\epsilon}\Gamma_{(\mu}\psi_{\nu)}\right),\nonumber\\
\delta A^I_{\mu}&=&\Re\left(\ft12\bar{\epsilon}\left(\partial_x X^I
\Gamma_{\mu}\lambda^{x}-iX^I\psi_{\mu}\right)\right),\nonumber\\
\delta X_I &=&\Re\left(\ft{i}{2}\partial_x
X_I\bar{\epsilon}\lambda^x\right),
\label{eq:bsusy}
\eea
for the bosons.  Here, ${\cal D}_\mu=\nabla_\mu-\fft32igV_IA^I_\mu$ is
the $U(1)$ covariant derivative.  Our notation here is that all spinors
are five-dimensional Dirac spinors, and the Dirac matrices satisfy the
Clifford algebra $\{\Gamma^\mu,\Gamma^\nu\}=2g^{\mu\nu}$.

\subsection{BPS black holes}

The $1/2$~BPS black hole solutions to gauged ${\cal N}=2$ supergravity
were obtained in \cite{BCS}, and have the form
\bea
ds^2 &=& -e^{-4U}f^2 dt^2 +
e^{2U}\left(f^{-2}dr^2+r^2d\Omega_3^2\right),\qquad
f^2 \equiv 1+g^2r^2e^{6U},\nonumber \\
A^I &=& e^{-2U}X^I dt,\qquad
X_I=\ft13e^{-2U}H_I,
\label{hole}
\eea
where
\begin{equation}
H_I=3V_I+\frac{q_I}{r^2}
\label{eq:harm}
\end{equation}
are a set of `harmonic' functions with constant electric charges $q_I$.
Note that the function $U(r)$ is determined implicitly through the
very special geometry constraint ${\cal V}=1$ where ${\cal V}$ is given
in (\ref{eq:calv}).

The solution in (\ref{hole}) is a $1/2$~BPS solution, and was constructed
by solving the Killing spinor equations $\delta\psi_\mu=0$ and
$\delta\lambda_x=0$ arising from (\ref{eq:fsusy}).  For the above background,
these equations take on the form
\begin{eqnarray}
\delta\psi_t&=&\left[\partial_t-ig+(-2ie^{-3U}fU'\Gamma_1
+gf(1+rU')\Gamma_0)P_{++}\right]\epsilon,\nonumber\\
\delta\psi_r&=&\left[\partial_r+U'-2U'P_{++}
+\ft12ge^{3U}f^{-1}(1+3rU')\Gamma_1\right]\epsilon,\nonumber\\
\delta\psi_\alpha&=&\left[(\hat\nabla_\alpha+\ft{i}2\Gamma_{01}
\hat\Gamma_\alpha)
-f(1+rU')\Gamma_1\hat\Gamma_\alpha P_{++}\right]\epsilon,\nonumber\\
\delta\lambda^x&=&-i\partial_r\phi^xe^{-U}f\Gamma_1P_{++}\epsilon,
\label{eq:bhkse}
\end{eqnarray}
where primes denote derivatives with respect to $r$, and numerical indices
$0,1$ denote frame indices with the obvious vielbeins
\begin{equation}
e^0=e^{-2U}f\,dt,\qquad e^1=e^Uf^{-1}dr.
\end{equation}
Angular coordinates on $S^3$ are given by $\alpha,\beta,\ldots$, and
the carets appearing in $\delta\psi_\alpha$ denote objects defined on
the {\it unit} sphere.
Here we have also introduced a family of projection operators
\be
\label{projections}
P_{\eta\tilde\eta}=\ft12[1+f^{-1}(\eta i\Gamma_0+\tilde\eta
gr e^{3U}\Gamma_1)],
\ee
where $\eta$ and $\tilde\eta$ are independently $\pm1$ (which we denote
$\pm$ in shorthand notation).  Although only $P_{++}$ shows up explicitly
in (\ref{eq:bhkse}), we will make use of the other signs of
$P_{\eta\tilde\eta}$ in future sections.

The Killing spinors corresponding to (\ref{eq:bhkse}) were constructed
in \cite{BCS}, and have the form
\bea
\epsilon_{++}&=& e^{igt}e^{-U}e^{-\fft{i}2 \Gamma_{012} \theta}e^{\fft12
  \Gamma_{23} \phi}e^{-\fft{i}2 \Gamma_{014} \psi}
\left(\sqrt{f+1}-\sqrt{f-1}\Gamma_1\right)\left(1-i\Gamma_0\right)\epsilon_0,
\label{eq:kspp}
\eea
where $\epsilon_0$ is an arbitrary constant spinor.  By construction,
$\epsilon_{++}$ satisfies the projection $P_{++}\epsilon_{++}=0$.  Here
we have used an explicit parameterization of the unit $S^3$ given by
\begin{equation}
d\Omega_3^2=d\theta^2+\sin^2\theta\,d\phi^2+\cos^2\theta\,d\psi^2,
\label{eq:units3}
\end{equation}
and have used $2,3,4$ to denote frame indices on $S^3$, with
\begin{equation}
\hat e^2=d\theta,\qquad \hat e^3=\sin\theta\,d\phi,\qquad
\hat e^4=\cos\theta\,d\psi.
\end{equation}
It is apparent from (\ref{eq:kspp}) that the Killing spinors split into
two parts: one related to the $t$--$r$ directions and the other corresponding
to Killing spinors on $S^3$.  This feature may be made explicit by choosing
a Dirac decomposition
\begin{equation}
\Gamma_0=i\sigma_2\times 1,\qquad\Gamma_1=\sigma_1\times1,\qquad
\Gamma_\alpha=\sigma_3\times\sigma_\alpha
\label{eq:diracd}
\end{equation}
along with the split $\epsilon=\varepsilon\times\eta$.  Here it is important
to realize that $\varepsilon$ may be taken to be Majorana (real) in the
$1+1$ dimensional space spanned by $t$ and $r$.

We now note that Killing spinors on $S^3$ corresponding to solutions of
\begin{equation}
(\hat\nabla_\alpha\pm\ft{i}2\hat\sigma_\alpha)\eta_\pm=0
\end{equation}
may be written explicitly as
\begin{equation}
\eta_\pm=e^{\mp\fft{i}2\sigma_1\theta}e^{\fft{i}2\sigma_3(\phi\mp\psi)}.
\label{eq:kss3}
\end{equation}
In this case, the projection (\ref{projections}) becomes
\begin{equation}
P_{\eta\tilde\eta}=\ft12[1+f^{-1}(-\eta\sigma_2+\tilde\eta
gre^{3U}\sigma_1)],
\end{equation}
while the Killing spinors (\ref{eq:kspp}) have the form
\begin{equation}
\epsilon_{++}=e^{igt}e^{-U}\left(\sqrt{f+1}-\sqrt{f-1}\sigma_1\right)
(1+\sigma_2)\varepsilon_0\times\eta_+.
\label{eq:kssplit}
\end{equation}
When working with supersymmetry, we will make use of both representations
(\ref{eq:kspp}) and (\ref{eq:kssplit}) interchangeably, whenever convenient.

\section{\label{sec:vary} Fermion zero modes and black hole superpartners}

Before proceeding to analyze the fermion zero modes, we find it useful
to familiarize ourselves with the form of Killing spinors in AdS$_5$.
Thus we first analyze the complete set of Killing spinors on the
maximally supersymmetric AdS space, and then demonstrate that half
of the original AdS Killing spinors naturally map into Killing spinors
in the presence of the black holes, while the other half become fermion
zero modes.

In general, of course, any spinor that does not solve the Killing spinor
equations $\delta\psi_\mu=0$, $\delta\lambda_x=0$ may be considered to
be zero modes.  However, one has to be careful in identifying physically
distinct configurations, as opposed to pure supergauge degrees of freedom.
The importance here is the recognition of the global part of the supersymmetry
algebra as the representation generating part.  In this sense, we demand
that the fermion zero modes are explicitly constructed to solve an
alternate projection $P_{-+}$, distinct from $P_{++}$ of the Killing
spinors.  Note, however, that $P_{-+}$ is not the complement of $P_{++}$;
that is reserved for $P_{--}$ satisfying $P_{++}+P_{--}=1$.  It instead
defines orthogonality with respect to the Dirac inner product,
$\overline{P_{(-\eta) (\tilde{\eta})}\epsilon_1}P_{\eta
\tilde{\eta}}\epsilon_2=0$.
This is because $\bar{\epsilon}=\epsilon^{\dagger}\Gamma_0$
and so the $\tilde{\eta}$ term changes sign when permuted past $\Gamma_0$.
Because of the background AdS curvature, this situation is
somewhat different from that used in \cite{Aichelburg:1986wv}, where
in addition to satisfying a supergauge choice $\gamma^\mu\delta\psi_\mu=0$
the fermion zero modes also satisfied the complementary projection
$P_-$ instead of $P_+$.

\subsection{Supersymmetry in $AdS_5$}

To highlight the above issues, we now consider supersymmetry in the $AdS_5$
vacuum.  This is readily obtained by taking $U=0$ and $q_I=0$ (so that
$X_I=V_I$ are constants) in the black hole ansatz of (\ref{hole}).  In
this case, the gaugino variation trivially vanishes, and (\ref{eq:bhkse})
reduces to the set
\bea
\delta \psi_{t} &=& \left[ \partial_t - (\ft32\mp\ft12) ig
+gf\Gamma_0 P_{\pm +}\right]\epsilon, \nonumber\\
\delta \psi_{r} &=& \left[ \partial_r + \ft12gf^{-1}\Gamma_1\right]\epsilon,
\nonumber\\
\delta \psi_{\alpha} &=& \left[\hat{\nabla}_{\alpha} \pm \ft{i}2
\Gamma_{01}\hat\Gamma_\alpha - f\Gamma_1\hat\Gamma_{\alpha}P_{\pm+}\right]
\epsilon.
\label{gtinos}
\eea
Unlike (\ref{eq:bhkse}), here we have made use of the identity
$P_{++}=P_{-+}+if^{-1}\Gamma_0$ to write the Killing spinor equations
using both types of projections.
One may solve these equations by starting with the solution $\epsilon_{++}$
of the last section with $U=0$.  To generate the solution to the $-+$
equation, simply note that to change $P_{++}$ into ${P_{-+}}$ one needs
to permute through a $\Gamma_1$.  This also leaves the $\psi_r$ equation
unchanged.  This implies that a spinor of the form
$\exp(igt)\Gamma_1\epsilon_{++}$ solves the $-+$ equations.  Pushing the
$\Gamma_1$ through until it is next to $\epsilon_0$, and then replacing
$\Gamma_1\epsilon_0\rightarrow \epsilon_0$ (since $\Gamma_1 \epsilon_0$ is
just as arbitrary as $\epsilon_0$) gives the other solution.  The AdS$_5$
Killing spinors may then be written as.
\be
\epsilon_{\pm +}=e^{i(\fft32\mp\fft12)gt}e^{\mp\frac{i}{2}\Gamma_{012}\theta}
e^{\fft12 \Gamma_{23}\phi}e^{\mp
  \frac{i}{2}\Gamma_{014}\psi}
\left(\sqrt{f+1}-\sqrt{f-1}\Gamma_1\right)
\left(1\mp i\Gamma_0\right)\epsilon_0,
\ee
where $\epsilon_{0}$ is again an arbitrary constant spinor.
The sign of $i\Gamma_0$ is directly
connected to the projection appearing next to $\epsilon_0$ giving that we
can replace the $\mp$ sign in the exponential with an $+i\Gamma_0$.  This
results in the usual form for the $AdS_5$ Killing spinors except for
the extra factor of $\exp(\fft32 igt)$. This extra factor is a direct
result of the gauge choice for $A^I$, but is otherwise physically
insignificant.

It ought to be apparent that, taken together, the complete
set of Killing spinors, $\epsilon_{++}$ and $\epsilon_{-+}$,
guarantee the maximal supersymmetry of the AdS$_5$ background.  As
seen from the $\delta\psi_\alpha$ equation of (\ref{gtinos}), the AdS$_5$
Killing spinors have a natural realization in terms of both types of
Killing spinors on $S^3$, namely $\eta_+$ and $\eta_-$ of (\ref{eq:kss3}).
using the standard Dirac decomposition (\ref{eq:diracd}), the above
Killing spinors take on the form
\begin{equation}
\label{dirsplit}
\epsilon_{\pm+}=e^{i(\fft32\mp\fft12)gt}
\left(\sqrt{f+1}-\sqrt{f-1}\sigma_1\right)
(1\pm\sigma_2)\varepsilon_0\times\eta_\pm.
\end{equation}
By construction, the above spinors $\epsilon_{\pm +}$ satisfy the projections
\be
\label{projection}
P_{\pm +}\epsilon_{\pm +}=0.
\ee
The $P_{++}$ case gives pure AdS$_5$ spinors which, when multiplied by
$e^{-U}$ and with appropriate modification to $f(r)$, correspond to the
preserved black hole supersymmetries identified in the last section.
We should also note that the pure AdS$_5$ $++$ spinors match the black
hole $++$ spinors when $r\rightarrow \infty$ (so that $U\rightarrow 0$)
because the space becomes asymptotically AdS$_5$.

The ${-+}$ solutions for pure AdS$_5$, on the other hand, are broken
supersymmetries when generalized to the black hole solution; they
correspond to fermion zero modes in this background.
Although any spinor not satisfying the $P_{++}$ projection would be
sufficient to realize the fermion zero mode algebra, the
$\epsilon_{-+}$ are particularly convenient because they reduce
to standard Killing spinors in an asymptotically AdS$_5$ spacetime
and hence represent genuine fermion zero modes related to the black
hole geometry (as opposed to supergauge transformations of pure AdS$_5$).
For convenience, we will drop the label $-+$ from zero mode spinors in
future sections.  Thus
\bea
\epsilon\equiv\epsilon_{-+}&=&e^{2igt}e^{\alpha U}
e^{\frac{i}{2}\Gamma_{012}\theta}
e^{\fft12 \Gamma_{23}\phi}e^{\frac{i}{2}\Gamma_{014}\psi}
\left(\sqrt{f+1}-\sqrt{f-1}\Gamma_1\right)
\left(1+ i\Gamma_0\right)\epsilon_0\nonumber\\
&=&e^{2igt}e^{\alpha U}\left(\sqrt{f+1}-\sqrt{f-1}\sigma_1\right)
\left(1-\sigma_2\right)\varepsilon_0\times\eta_-.
\label{epsilon}
\eea
Here $\alpha$ is an arbitrary constant related to choice of supergauge
condition; it will drop out in all physical quantities below.
These properly identified fermion zero mode spinors will be the starting
point for the generation of the superpartners of the black hole solution
of \cite{BCS}.

\subsection{Zero Mode Identities}

The black hole superpartners will be obtained via (\ref{eq:deltadelta})
up to second order in the supersymmetry transformation $\delta$.  As
a result, we are often faced with the task of simplifying bilinear
expressions in fermion zero mode spinors of the form $(\bar\epsilon
\Gamma_{\cdots}\epsilon)$, where $\epsilon$ is given by (\ref{epsilon}).
Using the projection properties (\ref{projection}) for the zero mode
spinors, as well as Dirac conjugation, $\bar\epsilon=\epsilon^\dagger
\Gamma^0$, one may obtain several useful identities:
\bea
&&(\bar{\epsilon}\Gamma_1\epsilon) = 0,\kern3.2em
  (\bar{\epsilon}\Gamma_0\epsilon) = -if(\bar{\epsilon}\epsilon),\kern5.1em
  (\bar{\epsilon}\Gamma_{01}\epsilon) = igre^{3U}(\bar{\epsilon}\epsilon),
\nonumber\\
&&(\bar{\epsilon}\Gamma_0\hat\Gamma_{\alpha}\epsilon) = 0,\qquad
  (\bar{\epsilon}\Gamma_1\hat\Gamma_{\alpha}\epsilon) =
    \frac{f}{gre^{3U}}(\bar{\epsilon}\hat\Gamma_{\alpha}\epsilon),\qquad
  (\bar{\epsilon}\Gamma_{01}\hat\Gamma_{\alpha}\epsilon)=
    -\frac{i}{gre^{3U}}(\bar{\epsilon}\hat\Gamma_{\alpha}\epsilon).
\label{id}
\eea
We will make use of these identities below.

\subsection{Black hole superpartners}

To generate black hole superpartners, we will consider fermion zero mode
transformations up to second order in $\epsilon$ starting from the bosonic
background (\ref{hole}) constructed in \cite{BCS}.  The first order
variations using the zero modes will generate a fermionic (gravitino and
gaugino) background.  Rewriting (\ref{eq:bhkse}) with the substitution
$P_{++}=P_{-+}+if^{-1}\Gamma_0$, and noting from (\ref{projection})
that $P_{-+}\epsilon=0$ for a fermion zero mode $\epsilon$, we obtain
\bea
\delta \psi_t &=& -U'\left[2e^{-3U}\Gamma_{01} + igr\right]\epsilon,
\nonumber \\
\delta \psi_r &=& U'\left[\alpha+1-2if^{-1}\Gamma_0\right]\epsilon,
\nonumber \\
\delta \psi_{\alpha} &=& -irU'\Gamma_{01}\hat\Gamma_{\alpha}\epsilon,
\nonumber\\
\delta\lambda^x &=& -e^{-U}\partial_r\phi^x\Gamma_{01}\epsilon.
\label{delpsilam}
\eea
Note that $\hat\Gamma_\alpha$ are Dirac matrices on the unit sphere, and
are related to the full five-dimensional matrices by
$\Gamma_\alpha=re^U\hat\Gamma_\alpha$.

We now turn to the terms second order in the supersymmetry variation, where
the bosonic fields receive corrections.  To obtain the second order
variations $\delta\delta(\hbox{boson})$, we may simply take their first
variations in (\ref{eq:bsusy}), and replace the fermions with their
first variations given above in (\ref{delpsilam}).  All other
contributions would be set to zero when evaluated for a bosonic
background.  Using the identities (\ref{id}) we find that the non-zero
variations of the metric are
\bea
\delta\delta
g_{tt}&=&-grU'f^2e^{-2U}\Re\,(\bar{\epsilon}\epsilon)=
          8grU'e^{2(\alpha+1)U}g_{tt}(\bar{\varepsilon}_0\varepsilon_0)N,
\nonumber\\
\delta\delta g_{t\alpha}&=&
   \ft32 (gr)^{-1}U'e^{-6U}
   \Re\,\left(i\bar{\epsilon}\Gamma_{\alpha}\epsilon\right)
   =-12rU'e^{2(\alpha-1)U}(\bar{\varepsilon}_0\varepsilon_0)\hat{K}_{\alpha},
\nonumber \\
\delta\delta g_{rr}&=&
   -2grU'f^{-2}e^{4U}\Re(\bar{\epsilon}\epsilon) =
      -16grU'e^{2(\alpha+1)U}g_{rr}(\bar{\varepsilon}_0\varepsilon_0)N,
\nonumber\\
\delta\delta g_{\alpha \beta} &=&
   grU' e^{2U} g_{\alpha \beta}
   \Re\,(\bar{\epsilon}\epsilon)=8grU'e^{2(\alpha+1)U}g_{\alpha\beta}
   (\bar\varepsilon_0\varepsilon_0)N.
\label{varg}
\eea
To obtain the final expressions on each line, we have decomposed the
fermion zero mode spinors according to (\ref{epsilon}) and taken
$\sigma_2 \varepsilon_0=-\varepsilon_0$ to satisfy
the projection $(1-\sigma_2)$ in (\ref{epsilon}) for the zero mode
spinors.  We have also defined
\begin{equation}
N=(\eta^\dagger_-\eta_-),\qquad
\hat K_\alpha=(\eta^\dagger_-\hat\sigma_\alpha\eta_-).
\label{eq:etaks}
\end{equation}
Here, $\hat K_\alpha$ is a Killing vector on the unit $S^3$, and the
decomposition (\ref{epsilon}) yields the relation
$(i\bar\epsilon\Gamma_\alpha\epsilon)=-8gr^2e^{2(\alpha+2)U}(\bar
\varepsilon_0\varepsilon_0)\hat K_\alpha$.  In addition, the non-trivial
double supersymmetry variations of the matter fields are
\bea
\delta\delta A^I_t &=& -\ft32 grU' X^I
   \Re\,(\bar{\epsilon}\epsilon)=-12grU'e^{2\alpha U}X^I(\bar\varepsilon_0
   \varepsilon_0)N,\nonumber \\
\delta \delta A^I_{\alpha} &=&\ft12(gr)^{-1}
   \left(U'X^I+\partial_rX^I\right)e^{-4U}
   \Re\,\left(i\bar{\epsilon}\Gamma_{\alpha}\epsilon\right)
   =-4r\left(U'X^I+\partial_rX^I\right)e^{2\alpha U}
   (\bar{\varepsilon}_0\varepsilon_0)\hat{K}_{\alpha},\qquad
\eea
for the gauge fields, and
\be
\delta\delta X_I =\ft12gr\partial_rX_Ie^{2U}
                    \Re\,(\bar{\epsilon}\epsilon)
=-4gr\partial_rX_Ie^{2(\alpha+1)U}(\bar\varepsilon_0\varepsilon_0)N,
\ee
for the scalars.

While the exponential factor $\exp(\alpha U)$ in (\ref{epsilon}) appears
in the above expressions, this factor goes to unity asymptotically as
$r\to\infty$.  Since $\alpha$ enters nowhere else, the actual value of
of $\alpha$ is unphysical.  For convenience, we take $\alpha=-1$ and
furthermore define the spinor bilinear
\begin{equation}
\lambda\equiv4(\bar\varepsilon_0\varepsilon_0).
\label{eq:sbilin}
\end{equation}
Using the expression (\ref{eq:deltadelta}) for a finite supersymmetry
transformation, we now observe that, up to second order in the
supersymmetry variation ({\it i.e.}~to lowest order in $\lambda$),
the bosonic fields may be expressed as
\bea
ds_{(\mathrm{tot})}^2&=&-e^{-4U}f^2(1+grU'\lambda N)dt^2
+e^{2U}[f^{-2}(1-2grU'\lambda N)dr^2
+r^2(1+grU'\lambda N)d\Omega_3^2]\nonumber\\
&&\qquad-3rU'e^{-4U}\lambda\,dt\hat K,\nonumber\\
A_{(\mathrm{tot})}^I&=&e^{-2U}X^I(1-\ft32 grU'\lambda N)dt -
\ft12r\left(U'X^I+\partial_r X^I\right)
   e^{-2U}\lambda\,\hat{K},\nonumber\\
X_I^{(\mathrm{tot})}&=&X_I-\ft12 gr\partial_r X_I\lambda N.
\label{eq:spartners}
\eea
Here, $\hat K=\hat K_\alpha d\theta^\alpha$ is the $1$-form associated
with the Killing vector $\hat K^\alpha\fft\partial{\partial\theta^\alpha}$.
where the $\alpha$ index is raised and lowered using the metric on the
{\it unit} sphere, (\ref{eq:units3}).  In the following section, we will
examine the superpartner solutions (\ref{eq:spartners}), and in particular
extract the superpartner shifts to the energy (mass), angular momentum and
$R$-charge of the original black hole.

\section{\label{sec:charges} Properties of the black hole superpartners}

Having constructed a set of black hole superpartners, (\ref{eq:spartners}),
in the ${\cal N}=2$ theory, we now set out to explore their properties.
We start by observing from (\ref{eq:spartners}) that angular momentum
(spin) is generated for the superpartners because of the off-diagonal
metric component proportional to $dt\hat K$.  This is of course expected,
as from a semi-classical point of view we expect the superpartners of the
spinless black hole to carry precisely spin-1/2.  We also see that the
effective Newtonian potential in $g_{tt}$ is shifted by a multiplicative
factor $(1+grU'\lambda N)$.  It is this shift that indicates that the
superpartner energies no longer coincide with that of the original solution.
This is a feature of supersymmetry in AdS spacetimes, and the energy shift
clearly vanishes in the Minkowski limit $g\to0$.

\subsection{Energy and angular momentum}

In order to make these observations on energy and angular momentum more
precise, we make use of holographic renormalization in AdS and in
particular the boundary stress tensor method for defining asymptotically
conserved quantities
\cite{Henningson:1998gx,Balasubramanian:1999re,Emparan:1999pm}.
Given a gravitational action $I[g_{\mu\nu}]$, the boundary stress
tensor is simply \cite{Brown:1992br}
\begin{equation}
T^{ab}=\fft2{\sqrt{-h}}\fft{\delta I}{\delta h_{ab}}
=-\fft1{8\pi G_5}(\Theta^{ab}-\Theta h^{ab}),
\label{eq:tunreg}
\end{equation}
where $h_{ab}$ is the boundary metric, $h_{\mu\nu}=g_{\mu\nu}-n_\mu
n_\nu$, with $n_\mu$ a unit normal to the boundary.  In addition,
$\Theta^{ab}$ is the extrinsic curvature tensor, which may be expressed as
\begin{equation}
\Theta_{ab}=-\fft12\left(\nabla_a n_b + \nabla_b n_a - n_a
n^c\nabla_c n_b - n_b n^c\nabla_c n_a\right).
\end{equation}
Note that the covariant derivatives are with respect to the $5$-dimensional
metric.  We have also used $a,b,c,\ldots$ to denote indices on the
boundary.  The expression (\ref{eq:tunreg}), while divergent, may
be regulated via the addition of boundary counterterms.  For this particular
situation, the appropriate counterterms were determined using the
Hamilton-Jacobi method in \cite{counter}.  The resulting renormalized
stress tensor is given by
\be
8 \pi G_5 T_{ab}  =
-\left(\Theta_{ab}-\Theta h_{ab}\right)\nonumber \\
               +\left(W(\phi)h_{ab}-\frac{1}{4g}\left(2 {\mathcal
  R}(h)_{ab}- {\mathcal R}(h)h_{ab}\right)\right).
\label{T}
\ee
Here $W(\phi)$ is the superpotential given in (\ref{eq:spot}) and
$\mathcal R_{ab}$ is the Ricci curvature on the boundary.
We note that although there are fermions present in the superpartner
configuration, (\ref{eq:spartners}), their behavior in the stress tensor
is dominated at large $r$
by a factor of $(U')^2$, and so the fermions fall off too
fast at the boundary to contribute to (\ref{T}).

To explore the boundary stress tensor, we take the black hole solution
of (\ref{eq:spartners}) and expand near the boundary at $r\to\infty$.
Using $r$ as the natural radial direction, the unit normal vector
$n_\mu$ has as its only non-vanishing component
$n_r=e^Uf^{-1}(1-grU'\lambda N)$, where we are only concerned with the
lowest order in $\lambda$.  Corresponding to this normal direction,
the four-dimensional constant $r$ surfaces are given by
\bea
ds_4^2&\equiv&h_{ab}dx^a dx^b\nonumber\\
&=&-e^{-4U}f^2(1+grU'\lambda N)dt^2 \nonumber \\
&&+e^{2U}r^2 (1+grU'\lambda N) \hat{g}_{\alpha\beta}
\left(d\theta^{\alpha}+\ft32r^{-1}U'e^{-6U}\lambda\hat{K}^{\alpha}dt\right)
\left(d\theta^{\beta}+\ft32r^{-1}U'e^{-6U}\lambda\hat{K}^{\beta}dt\right),
\qquad
\label{eq:bmet}
\eea
where we have only worked to linear order in $\lambda$.  Note that we
have further chosen an ADM-like foliation of the boundary metric, with
shift vectors related to angular momentum.  Furthermore, given the
unit normal, it is straightforward to compute the extrinsic curvature
tensor from the four-dimensional metric:
\bea
\Theta_{ab}&=&-\ft12(\nabla_a n_b + \nabla_b n_a)=-\ft12n^r\partial_rh_{ab}
=-\ft12e^{-U}f\left(1+grU'\lambda N\right)\partial_rh_{ab}.
\label{eq:theta}
\eea

To compute the counterterm contributions in (\ref{T}), we also need the
superpotential and the Ricci curvature of the boundary metric
(\ref{eq:bmet}).  Since $r$ may be taken as constant with respect to
the four-dimensional metric, its intrinsic curvature has a simple form,
with only the $S^3$ being curved.  In other words, we have
\be
\mathcal R_{tt}=0,\qquad\mathcal R_{t\alpha}=0,\qquad
\mathcal R_{\alpha\beta}=2r^{-2}e^{-2U}(1-grU'\lambda N)h_{\alpha \beta}.
\label{eq:calr}
\ee
In addition, using (\ref{eq:spot}) and the form of $X_I^{(\mathrm{tot})}$
given in (\ref{eq:spartners}), the superpotential has the form
\begin{equation}
W=3gV_IX^I=3g(1-\ft12gr\lambda N\partial_r)(V_IX^I)
=3g(1-\ft12gr\lambda N\partial_r)[e^{2U}(1+rU')].
\label{eq:w}
\end{equation}

To proceed, we now need to specify the functional form of $U(r)$.
Although in some cases (such as the STU model) a closed-form expression
may be given for $U$, here it is sufficient for us to assume that $U$ has
an expansion in inverse powers of $r^2$ of the form
\be
U=\frac{\alpha_1}{r^2}+\frac{\alpha_2}{r^4}+\cdots.
\ee
In this case, $\Theta_{ab}$, $\mathcal R_{ab}$ and $W$ given in
(\ref{eq:theta}), (\ref{eq:calr}) and (\ref{eq:w}) may be expanded for
large $r$ and inserted into (\ref{T}).  We find, to lowest non-trivial
order
\bea
8\pi G_5T_{tt}&=&-\frac{g}{r^2}\left(6\alpha_1(1+\ft12g\lambda N)+
\frac{3}{8g^2} \right),\nonumber\\
8\pi G_5 T_{t \alpha} &=&-\fft{g}{r^2}\left(
6\alpha_1\lambda\hat{K}_{\alpha}\right).
\eea
Note that there is also a contribution to $T_{\alpha\beta}$, which we
have not computed (since it has no r\^ole to play in extracting conserved
charges).

We must now extract the energy and angular momentum from the above
expressions for the boundary stress tensor.  To obtain the charge
$Q_\xi$ associated with a Killing vector $\xi$, we take
\be
Q_{\xi}=-\lim_{r\to\infty}
\int d^3\theta \sqrt{\gamma}(u^{a}T_{ab}\xi^b),
\label{Q}
\ee
where $u$ defines the unit normal time direction, and $\gamma$ is the
induced metric for a constant time slice.  For the metric (\ref{eq:bmet}),
this expression takes the form
\be
Q_\xi=-\lim_{r\to\infty} \fft{\omega_3r^2}g(T_{ta}\xi^a),
\ee
where $\omega_3=2\pi^2$ is the volume of the unit $3$-sphere.
In particular, conjugate to the Killing vector $\xi=\fft\partial{\partial t}$,
we obtain the energy
\be
\label{mass}
E=-\lim_{r\to\infty}\fft{2\pi^2r^2}gT_{tt}
=\frac\pi{4G_5}\left(6\alpha_1(1+\ft12g\lambda N)
+\frac{3}{8g^2}\right).
\ee

To obtain the angular momentum, we must consider some properties of
Killing vectors on the unit $3$-sphere.  In general, the unit $S^3$ admits
a set of $\mathrm{SO}(4)$ Killing vectors, which we may denote $\hat
K_\alpha^{(ij)}$, where $ij$ is an antisymmetric $\mathrm{SO}(4)$ index
pair.  These Killing vectors may be normalized according to
\be
\int d^3{\theta} \sqrt{\hat{g}}
\hat{K}^{(ij)}_{\alpha}\hat{K}^{(lm)\,\alpha}=
\ft12\left(\delta^{il}\delta^{jm}-\delta^{im}\delta^{jl}\right)\omega_3.
\ee
On the other hand, the Killing vector $\hat K^\alpha$ constructed from
Killing spinors in (\ref{eq:etaks}) is naturally given as an
$\mathrm{SU}(2)_-$ Killing vector, corresponding to the decomposition
$\mathrm{SO}(4)=\mathrm{SU}(2)_+\times\mathrm{SU}(2)_-$.  In particular,
Killing vectors corresponding to $\mathrm{SU}(2)_+$ and
$\mathrm{SU}(2)_-$ arise from $(\eta_+^\dagger\hat\sigma^\alpha\eta_+)$
and $(\eta_-^\dagger\hat\sigma^\alpha\eta_-)$, respectively.  While
$\hat K^\alpha\equiv(\eta_-^\dagger\hat\sigma^\alpha\eta_-)$ depends on
the explicit Killing spinor $\eta_-$, we may always choose coordinates
such that $\hat K^\alpha$ is aligned along the $T^3$ direction of
$\mathrm{SU}(2)$.  For the unit $S^3$ given in (\ref{eq:units3}), this
corresponds to taking
\begin{equation}
\hat K^\alpha=\fft\partial{\partial\phi}+\fft\partial{\partial\psi}
=\hat K^{(12)\,\alpha}+\hat K^{(34)\,\alpha},
\label{eq:kvect3}
\end{equation}
where we have identified $\phi$ and $\psi$ with rotations in the
1-2 and 3-4 planes, respectively.  This also agrees with the natural
embedding of $\mathrm{SU}(2)_\pm$ in $\mathrm{SO}(4)$.
Using these expressions in (\ref{Q}), we now read off the angular
momentum
\be
\label{angP}
J^{ij}=-\lim_{r\to\infty} \fft{r^2}g \int d^3\theta\sqrt{\hat g}
T_{t\alpha}\hat K^{(ij)\,\alpha}
=\fft{3\alpha_1\lambda}{4\pi G_5}\int d^3\theta\sqrt{\hat g}\hat K_\alpha
\hat{K}^{(ij)\,\alpha},
\ee
so that
\begin{equation}
J^{12}=J^{34}=\fft\pi{4G_5}(3\alpha_1\lambda).
\label{eq:angmom}
\end{equation}

We should note that, while these definitions for energy and angular
momentum were obtained for AdS black holes, they exactly match their
Minkowski black hole counterparts in the case where the black hole is
``small.''  When we say
that the black hole is small, we mean that all length scales associated
with it are small compared to the radius of AdS.  In such a case
there is a region of space such that $gr \ll \delta_{<},\;\;
\alpha_k/r^{2k}\ll \delta_{>},\;\; \delta_{<}\ll \delta_{>} \ll 1$.
The black hole's energy and angular momentum may then be read off from
the metric using a standard ADM prescription; these expressions should
furthermore agree with the above (up to the Casimir energy, which is
absent in the ADM mass).  In fact, we would have {\it chosen} the
definitions of the conserved charges in such a way as to have this
happen (by modifying them with multiplicative constants).
The fact that they do agree merely confirms that we have defined them
in an appropriate manner.

\subsection{The $R$-charge and magnetic dipole moment}

The conserved gauge charges are straightforward to obtain, and do
not require a counterterm prescription.  Based on the Maxwell equation
of motion from (\ref{action}), we obtain the conserved Noether
(electric) charges
\be
Q_I=\lim_{r\to\infty}\frac{1}{\omega_3}\int d^3\theta
\sqrt{-g}\,G_{IJ}F^{J\,rt}.
\ee
Now, we simply note that the first order corrections in
$\lambda$ to $\sqrt{-g}\,G_{IJ}$, when contracted with $F^{J\, rt}$,
fall off too fast to contribute.  The only modification to the
charge of the superpartners therefore comes from a direct shift
in $A$.  From (\ref{eq:spartners}), we obtain
\begin{equation}
Q_I=q_I-9V_I\alpha_1\lambda N,
\end{equation}
where $q_I$ were the original black hole electric charges, given
in (\ref{eq:harm}).

For BPS states, we are more specifically interested in the $R$-charge,
given as the electric charge of $\mathcal A_\mu=V_IA^I_\mu$.  In
this case, we may simply read off the $R$ charge from $V_IA_t^I$:
\be
V_IA^I_t=1-\frac{2\alpha_1}{r^2}(1-\ft32g\lambda N)+\cdots.
\ee
Identifying this expression with $Q/(2r^2)$ (up to the constant,
which is pure gauge), we obtain
\begin{equation}
Q=-4\alpha_1(1-\ft32g\lambda N).
\end{equation}
Finally, we may also read off the graviphoton magnetic dipole moment from
\be
\label{mag}V_IA^I_{\alpha}=\frac{\alpha_1\lambda}{r^2}\tilde{K}_{\alpha}=
\fft{\alpha_1\lambda}{r^2}(\hat K^{(12)}_\alpha+\hat K^{(34)}_\alpha).
\ee
Identifying this with $-\fft12\mu_{ij}\hat K_\alpha^{(ij)}r^{-2}$ yields
\begin{equation}
\mu_{12}=\mu_{34}=-\alpha_1\lambda,
\label{eq:magdip}
\end{equation}
for the magnetic dipole moment $\mu_{ij}$.  We will discuss the relation
between these charges in the next section.

\section{\label{sec:dis}Discussion}

Given the above construction of BPS black hole superpartners in AdS$_5$,
and the further determination of their conserved charges, we now
demonstrate that the structure of the superpartners is consistent with
representation theory.  In particular, we have worked in the context of
gauged $D=5$, ${\cal N}=2$ supergravity, with superalgebra
$\mathrm{SU}(2,2|1)$.  Recall that highest weight representations
\cite{Flato:1983te,Dobrev:1985qv} (see also Appendix~B of
\cite{Freedman:1999gp}) may be
labeled by $\mathcal D(E_0,j_1,j_2;r)$ where the lowest energy $E_0$,
spins $j_1$ and $j_2$, and $R$-charge $r$ label the compact bosonic
subalgebra
\begin{equation}
\mathrm{SU}(2,2|1)\supset\mathrm{SO}(2,4)\times\mathrm U(1)
\supset\mathrm{SO}(2)\times\mathrm{SU}(2)\times\mathrm{SU}(2)
\times\mathrm U(1).
\end{equation}
This superalgebra allows for two types of short multiplets (chiral and
non-chiral) in addition to ordinary long multiplets.  The long multiplets
generically contain $2^4=16$ states, while the short ones contain $2^2=4$
states.  The non-chiral multiplets are of the form
\begin{eqnarray}
\mathcal D(E_0=2j+1,j,j;0)&=&D(E_0,j,j)_0+D(E_0+\ft12,j+\ft12,j)_{-1}
\nonumber\\
&&\qquad +D(E_0+\ft12,j,j+\ft12)_1+D(E_0+1,j+\ft12,j+\ft12)_0,
\end{eqnarray}
while the chiral ones are
\begin{eqnarray}
\mathcal D(E_0=\ft32r,j,0,r)&=&D(E_0,j,0)_r+D(E_0+\ft12,j+\ft12,0)_{r-1}
\nonumber\\
&&\qquad +D(E_0+\ft12,j-\ft12,0)_{r-1}+D(E_0+1,j,0)_{r-2},\nonumber\\
{\cal D}(E_0=-\ft32r,0,j,r)&=&D(E_0,0,j)_r+D(E_0+\ft12,0,j+\ft12)_{r+1}
\nonumber\\
&&\qquad +D(E_0+\ft12,0,j-\ft12)_{r+1}+D(E_0+1,0,j)_{r+2}.
\label{eq:chiralshort}
\end{eqnarray}

Since the BPS black holes of \cite{BCS} carry non-zero $R$-charge,
they ought to correspond to the chiral short multiplet given above.
To see this, we identify the energy, angular momentum and $R$-charge
obtained in the previous section as
\begin{equation}
E=\fft{\pi}{4G_5}(6\alpha_1+3g\alpha_1\lambda N+\ft38g^{-2}),\qquad
J^{12}=J^{34}=\fft{\pi}{4G_5}3\alpha_1\lambda,\qquad
Q=-4\alpha_1+6g\alpha_1\lambda N,
\end{equation}
Removing the Casimir energy from $E$, and dropping the prefactor
$\pi/4G_5=\omega_3/8\pi G_5$ from gravitational quantities, we see
that the appropriate identification of $\mathrm{SU}(2,2|1)$ quantum
numbers is as follows:
\begin{equation}
E_0=6\alpha_1+3g\alpha_1\lambda N,\qquad j_1=0,\qquad
j_2=3\alpha_1\lambda,\qquad r=-4\alpha_1+6g\alpha_1\lambda N.
\label{eq:qnum}
\end{equation}
Setting $\lambda=0$ for the original bosonic solution then yields
\begin{equation}
D(E_0,0,0)_r,\qquad E_0=-\ft32r=6\alpha_1,
\end{equation}
corresponding to the lowest weight component of
$\mathcal{D}(E_0=-\ft32r,0,j=0,r)$ given in (\ref{eq:chiralshort}).

Turning now to the superpartners, we first note from (\ref{eq:kspp})
that $\varepsilon_0$ satisfies a projection $\sigma_2\varepsilon_0=
-\varepsilon_0$.  Hence this two-component Majorana spinor
in fact has only one independent real component, which may be taken
as an unimportant real multiplicative constant in the product
$\epsilon=\varepsilon_0\times\eta_-$.  In other words, the interesting
fermion zero mode algebra arises from the Killing spinors $\eta_-$ on
the sphere, and not from $\varepsilon_0$ itself.
Based on standard representation theory techniques, we see that this
algebra is essentially that of fermionic creation and annihilation
operators.  Thus we view the two-component Dirac spinor $\eta_-$
and its conjugate $\eta_-^\dagger$as a pair of creation and annihilation
operators
\begin{equation}
\eta_-^\dagger=\pmatrix{a_\uparrow^\dagger&a_\downarrow^\dagger},\qquad
\eta_-=\pmatrix{a_\uparrow\cr a_\downarrow},
\label{eq:aadagger}
\end{equation}
with corresponding number operator
\begin{equation}
N=(\eta_-^\dagger\eta_-)=a_\uparrow^\dagger a_\uparrow+a_\downarrow^\dagger
a_\downarrow=n_\uparrow+n_\downarrow.
\end{equation}
Of course, given the semi-classical analysis of the previous sections,
the normalization of these operators is not so obvious.  Fortunately,
we have just seen that the parameter $\lambda$, defined in (\ref{eq:sbilin})
as $\lambda=4(\bar\varepsilon_0\varepsilon_0)$, is an ordinary $c$-number.
Thus we simply assume that the black hole superpartners with
non-vanishing spin will be normalized so that $j_2=\fft12$ (actually the
third component of $j_2$).  This corresponds to setting $6\alpha_1\lambda=1$
in (\ref{eq:qnum}).
To be somewhat more precise, there are actually two independent sets of
creation and annihilation operators, as indicated in (\ref{eq:aadagger}).
In this case, the third component of angular momentum $j_2$ in
the second $\mathrm{SU}(2)$ may be either `spin up' or `spin down',
depending on the choice of Killing spinors $\eta_-$ used to construct
the Killing vector $\hat K^\alpha$.  In fact, from (\ref{eq:etaks}),
we may write down
\begin{equation}
T^3\equiv\ft12(\eta_-^\dagger\sigma^3\eta_-)=\ft12(a_\uparrow^\dagger
a_\uparrow-a_\downarrow^\dagger a_\downarrow)=\ft12(n_\uparrow-n_\downarrow).
\label{eq:t3def}
\end{equation}
We now see that the choice of Killing vector in (\ref{eq:kvect3}) is
overly restrictive.  As a consequence, instead of having $j_2=3\alpha_1
\lambda$ in (\ref{eq:qnum}), we ought to write $m_2=6\alpha_1\lambda T^3$
with $T^3$ given in (\ref{eq:t3def}), where $m_2$ is the third component
of $j_2$.

Given the above considerations, we see that the superpartner quantum numbers
read off from (\ref{eq:qnum}) fit the representations
\begin{equation}
D(E_0+\ft12gN,0,m_2=T^3)_{r+gN},\qquad E_0=-\ft32r,
\end{equation}
where $N=n_\uparrow+n_\downarrow$ and $T^3=\fft12(n_\uparrow-n_\downarrow)$.
Note here that the spin $j_2$ is given implicitly in terms of the angular
momentum representations $|j_2,m_2\rangle$.  Since the number operators
$n_\uparrow$ and $n_\downarrow$ independently take on the values
$0,1$, we identify precisely the $4$ states of the short multiplet.
In particular, we have $N=0,1,2$, with corresponding spins $j_2=0,\fft12,0$.
As the dimensionful quantities $E_0$ and $r$ are measured with respect to
the AdS inverse radius $g$, the above expression is in complete
agreement with the chiral short representation of (\ref{eq:chiralshort})
with superspin $j=0$.
Thus we have demonstrated that, in fact, working to second order in the
supersymmetry transformations is sufficient to reproduce the appropriate
zero mode algebra of the corresponding supersymmetry algebra.

\subsection{The gyromagnetic ratio}

Following \cite{Duff:1996bs,Duff:1998ef}, we may also compute the
gyromagnetic ratio of the black hole superpartners.  Here we make
use of the definition
\be
\mu^{ij}=\frac{\tilde{g}Q}{2M}J^{ij},
\label{eq:gyro}
\ee
where $\tilde g$ denotes the gyromagnetic ratio (to distinguish it from
the inverse AdS radius).
The magnetic dipole moment was identified in (\ref{eq:magdip}) to be
\begin{equation}
\mu_{12}=\mu_{34}=-\alpha_1\lambda,
\end{equation}
which is clearly proportional to the angular momentum $J_{12}=J_{34}
=(\pi/4G_4)3\alpha_1\lambda$ given in
(\ref{eq:angmom}).  To compute the gyromagnetic ratio, we further use
the mass and charge of the original bosonic solution,
$M=(\pi/4G_5)6\alpha_1$ and $Q=-4\alpha_1$ to obtain
\begin{equation}
(-\alpha_1\lambda)=\fft{\tilde g(-4\alpha_1)}
{2(\pi/4G_5)6\alpha_1}\left(\fft\pi{4G_5}3\alpha_1\lambda\right),
\ee
which yields $\tilde g=1$.  This agrees with the asymptotically
Minkowski case previously obtained in \cite{Herdeiro:2000ap}.
We use the mass and charge of the original
system because $J_{ij}$ and $\mu_{ij}$ were only calculated to leading
order.

This result appears somewhat surprising, as it was proven in
\cite{Ferrara:1992nm,Giannakis:1997iw} that unbroken supersymmetry
in four dimensions is sufficient to ensure that $\tilde g$ is {\it exactly}
equal to $2$ for the superspin-$0$ multiplet.  In this case, the situation
is somewhat different, as we are working in five dimensions and furthermore
with the anti-de Sitter superalgebra.  It turns out, however, that it is
not the AdS nature of the system that leads to $\tilde g\neq2$, but rather
just the simple fact that $\tilde g=1$ is natural in five dimensions, at
least for superpartners of non-rotating black holes \cite{Herdeiro:2000ap}.
While $\mathcal N=2$ supersymmetry shares many common features between
four and five dimensions, there are differences as well.  Consider, for
example, the minimal (ungauged) supergravity multiplet
$(g_{\mu\nu},\psi_\mu,A_\mu)$, with supersymmetry transformation
\begin{eqnarray}
\delta\psi_\mu&=&[\nabla_\mu+\ft{i}8(\gamma_\mu{}^{\nu\rho}
-2(D-3)\delta_\mu^\nu\gamma^\rho)F_{\nu\rho}]\epsilon,\nonumber\\
\delta g_{\mu\nu}&=&\bar\epsilon\gamma_{(\mu}\psi_{\nu)},\nonumber\\
\delta A_\mu&=&-\ft{i}{D-3}\bar\epsilon\psi_\mu,
\end{eqnarray}
normalized in either $D=4$ or $D=5$ according to $[\delta_1,\delta_2]
\Phi=\fft12(\bar\epsilon_1\gamma^\mu\epsilon_2)\partial_\mu
\Phi+\cdots$, with $\Phi$ any of the fields in the multiplet.  This
system admits BPS (Reissner-Nordstrom) black holes of the form
\begin{eqnarray}
&&ds^2=-\mathcal H^{-2}dt^2+\mathcal H^{2/(D-3)}d\vec y^2,\nonumber\\
&&A_{(1)}=\ft2{D-3}\mathcal H^{-1}dt.
\end{eqnarray}
For a spherically symmetrical black hole with harmonic function
$\mathcal H=1+q/r^{D-3}$, application of the techniques of
\cite{Aichelburg:1986wv,Duff:1996bs,Duff:1998ef} to generate superpartners
yields
\begin{equation}
\delta\delta g_{ti}=(\bar\epsilon i\gamma_i{}^j\epsilon)\hat x_j
\fft{(D-2)q}{2r^{D-2}},\qquad
\delta\delta A_i=(\bar\epsilon i\gamma_i{}^j\epsilon)\hat x_j\fft{q}{(D-3)
r^{D-2}}.
\end{equation}
where $\epsilon$ is a fermion zero mode spinor.  After extracting then
angular momentum and magnetic moment from these expressions, and inserting
them into (\ref{eq:gyro}), we obtain
\begin{equation}
\tilde g=\fft2{D-3}=\cases{2&for $D=4$,\cr1&for $D=5$.}
\end{equation}
So we see that $\tilde g=1$ is actually expected in five dimensions,
regardless of whether the background is AdS or Minkowski.  Noting that
$\tilde g=1$ is the natural value in both IIA theory in ten dimensions
and maximal supergravity in eight dimensions \cite{Duff:1998ef}, it
rather appears that $\tilde g=2$ is a unique feature of four dimensions.

Finally, we note that while we have focused on the stationary BPS
solutions of \cite{BCS}, they actually have singular horizons or naked
singularities in the context of the STU model.  While this is a rather
undesirable feature, our present analysis is unaffected by such
singularities, as we only depend on the asymptotics away from the
singularity.  It would, however, be worthwhile to extend the fermion
zero mode construction to the case of the recently constructed
supersymmetric AdS$_5$ black holes supported by rotation
\cite{Gutowski:2004ez,Gutowski:2004yv}.  In addition, while to our
knowledge this method has only been applied to the generation of
superpartners of particle-like representations, nothing prevents it
from being extended to more general backgrounds with partially
broken supersymmetry.  It would be of particular interest to examine
the fermion zero modes in the recently constructed $1/2$~BPS backgrounds
\cite{Lin:2004nb,Liu:2004ru}, and to explore the r\^ole they may play
in the AdS/CFT context.

\begin{acknowledgments}
This project had its origins in earlier discussions with M.~J.~Duff
on constructing AdS black hole superpartners following the completion of
\cite{Duff:1999gh}.
WS and JTL wish to thank the Khuri lab at the Rockefeller University for its
hospitality while part of this work was completed.
This work was supported in part by the US Department of Energy under
grant DE-FG02-95ER40899 and the National Science Foundation under grant
PHY-0313416.
\end{acknowledgments}



\begin{thebibliography}{99}

\bibitem{Aichelburg:1986wv}
P.~C.~Aichelburg and F.~Embacher,
{\sl The Exact Superpartners Of N=2 Supergravity Solitons},
Phys.\ Rev.\ D {\bf 34}, 3006 (1986).

\bibitem{Duff:1996bs}
M.~J.~Duff, J.~T.~Liu and J.~Rahmfeld,
{\sl Dipole moments of black holes and string states},
Nucl.\ Phys.\ B {\bf 494}, 161 (1997)
[arXiv:hep-th/9612015].

\bibitem{Duff:1998ef}
M.~J.~Duff, J.~T.~Liu and J.~Rahmfeld,
{\sl $g = 1$ for Dirichlet 0-branes},
Nucl.\ Phys.\ B {\bf 524}, 129 (1998)
[arXiv:hep-th/9801072].

\bibitem{Balasubramanian:1998za}
V.~Balasubramanian, D.~Kastor, J.~H.~Traschen and K.~Z.~Win,
{\sl The spin of the M2-brane and spin-spin interactions via probe techniques},
Phys.\ Rev.\ D {\bf 59}, 084007 (1999)
[arXiv:hep-th/9811037].

\bibitem{Henningson:1998gx}
M.~Henningson and K.~Skenderis,
{\sl The holographic Weyl anomaly},
JHEP {\bf 9807}, 023 (1998)
[arXiv:hep-th/9806087].

\bibitem{Balasubramanian:1999re}
V.~Balasubramanian and P.~Kraus,
{\sl A stress tensor for anti-de Sitter gravity},
Commun.\ Math.\ Phys.\  {\bf 208}, 413 (1999)
[arXiv:hep-th/9902121].

\bibitem{Emparan:1999pm}
R.~Emparan, C.~V.~Johnson and R.~C.~Myers,
{\sl Surface terms as counterterms in the AdS/CFT correspondence},
Phys.\ Rev.\ D {\bf 60}, 104001 (1999)
[arXiv:hep-th/9903238].

\bibitem{Liu:2004it}
J.~T.~Liu and W.~A.~Sabra,
{\sl Mass in anti-de Sitter spaces},
arXiv:hep-th/0405171.

\bibitem{counter}
A.~Batrachenko, J.~T.~Liu, R.~McNees, W.~A.~Sabra and W.~Y.~Wen,
{\sl Black hole mass and Hamilton-Jacobi counterterms},
arXiv:hep-th/0408205.

\bibitem{sugra}
M.~Gunaydin, G.~Sierra and P.~K.~Townsend,
{\sl The Geometry Of $N=2$ Maxwell-Einstein Supergravity And Jordan Algebras},
Nucl.\ Phys.\ B {\bf 242}, 244 (1984).

\bibitem{gauged}
M.~Gunaydin, G.~Sierra and P.~K.~Townsend,
{\sl Gauging The $D=5$ Maxwell-Einstein Supergravity Theories: More On Jordan
Algebras},
Nucl.\ Phys.\ B {\bf 253}, 573 (1985).

\bibitem{BCS}
K.~Behrndt, A.~H.~Chamseddine and W.~A.~Sabra,
{\sl BPS black holes in $N = 2$ five dimensional AdS supergravity},
Phys.\ Lett.\ B {\bf 442}, 97 (1998)
[arXiv:hep-th/9807187].

\bibitem{Behrndt:1998jd}
K.~Behrndt, M.~Cvetic and W.~A.~Sabra,
{\sl Non-extreme black holes of five dimensional $N = 2$ AdS supergravity},
Nucl.\ Phys.\ B {\bf 553}, 317 (1999)
[arXiv:hep-th/9810227].

\bibitem{Brown:1992br}
J.~D.~Brown and J.~W.~York,
{\sl Quasilocal energy and conserved charges derived from the gravitational
action},
Phys.\ Rev.\ D {\bf 47}, 1407 (1993).

\bibitem{Flato:1983te}
M.~Flato and C.~Fronsdal,
{\sl Representations Of Conformal Supersymmetry},
Lett.\ Math.\ Phys.\  {\bf 8}, 159 (1984).

\bibitem{Dobrev:1985qv}
V.~K.~Dobrev and V.~B.~Petkova,
{\sl All Positive Energy Unitary Irreducible Representations Of Extended
Conformal Supersymmetry},
Phys.\ Lett.\ B {\bf 162}, 127 (1985).

\bibitem{Freedman:1999gp}
D.~Z.~Freedman, S.~S.~Gubser, K.~Pilch and N.~P.~Warner,
{\sl Renormalization group flows from holography supersymmetry and a
$c$-theorem},
Adv.\ Theor.\ Math.\ Phys.\  {\bf 3}, 363 (1999)
[arXiv:hep-th/9904017].

\bibitem{Herdeiro:2000ap}
C.~A.~R.~Herdeiro,
{\sl Special properties of five dimensional BPS rotating black holes},
Nucl.\ Phys.\ B {\bf 582}, 363 (2000)
[arXiv:hep-th/0003063].

\bibitem{Ferrara:1992nm}
S.~Ferrara and M.~Porrati,
{\sl Supersymmetric sum rules on magnetic dipole moments of arbitrary spin
particles},
Phys.\ Lett.\ B {\bf 288}, 85 (1992).

\bibitem{Giannakis:1997iw}
I.~Giannakis and J.~T.~Liu,
{\sl $N = 2$ supersymmetry and dipole moments},
Phys.\ Rev.\ D {\bf 58}, 025009 (1998)
[arXiv:hep-th/9711173].

\bibitem{Gutowski:2004ez}
J.~B.~Gutowski and H.~S.~Reall,
{\sl Supersymmetric AdS$_5$ black holes},
JHEP {\bf 0402}, 006 (2004)
[arXiv:hep-th/0401042].

\bibitem{Gutowski:2004yv}
J.~B.~Gutowski and H.~S.~Reall,
{\sl General supersymmetric AdS$_5$ black holes},
JHEP {\bf 0404}, 048 (2004)
[arXiv:hep-th/0401129].

\bibitem{Lin:2004nb}
H.~Lin, O.~Lunin and J.~Maldacena,
{\sl Bubbling AdS space and 1/2 BPS geometries},
JHEP {\bf 0410}, 025 (2004)
[arXiv:hep-th/0409174].

\bibitem{Liu:2004ru}
J.~T.~Liu, D.~Vaman and W.~Y.~Wen,
{\sl Bubbling 1/4 BPS solutions in type IIB and supergravity reductions on
$S^n\times S^n$},
arXiv:hep-th/0412043.

\bibitem{Duff:1999gh}
M.~J.~Duff and J.~T.~Liu,
{\sl Anti-de Sitter black holes in gauged $N = 8$ supergravity},
Nucl.\ Phys.\ B {\bf 554}, 237 (1999)
[arXiv:hep-th/9901149].

\end{thebibliography}
\end{document}